# No evidence for modifications of gravity from galaxy motions on cosmological scales

Jian-hua He*[1], Luigi Guzzo[2,3,4], Baojiu Li[1], Carlton M. Baugh[1]

[1] *Institute for Computational Cosmology, Department of Physics, Durham University, Durham DH1 3LE, UK*

[2] *Dipartimento di Fisica, Universita' degli Studi di Milano, via G. Celoria 16, 20133 Milano, Italy*

[3] *INAF – Osservatorio Astronomico di Brera, Via Brera 28, Milano, Italy*

[4] *INFN – Sezione di Milano, Via G. Celoria 16, 20133, Milano, Italy*

The recent discovery of gravitational waves [1]-[2] marks the culmination of a sequence of successful tests of the general theory of relativity (GR) since its formulation in 1915. Yet these tests remain confined to the scale of stellar systems or the strong gravity regime. A departure from GR on larger, cosmological scales has been advocated by the proponents of "modified" gravity theories [3] as an alternative to the *Cosmological Constant* $\Lambda$ [4] to account for the observed cosmic expansion history [5]-[6]. While indistinguishable in these terms by construction, such models on the other hand yield distinct values for the linear growth rate of density perturbations and, as a consequence, for the associated galaxy peculiar velocity field. Measurements of the resulting anisotropy of galaxy clustering, when spectroscopic *redshifts* are used to derive distances [7]-[8], have thus been proposed as a powerful probe of the validity of GR on cosmological scales [9]. However, despite significant effort in modelling such *redshift space distortions* (e.g. [10]), systematic errors remain comparable to current statistical uncertainties (e.g. [11]). Here, we present the results of a different forward-modelling approach, which fully exploits the sensitivity of the galaxy velocity field to modifications of GR. We use state-of-the-art, high-resolution N-body simulations of a standard GR ($\Lambda$CDM) [12] and a **compelling $f(R)$ model [13] – one of GR's simplest variants – to build simulated catalogues of stellar-mass-selected galaxies through a robust match to the Sloan Digital Sky Survey observations [14]. We find that, well within the uncertainties of this technique, $f(R)$ fails to reproduce the observed redshift-space clustering on scales 1-10 Mpc/h. Instead, the standard ΛCDM GR model agrees impressively well with the data. This result provides a strong confirmation, on cosmological scales, of the robustness of Einstein's general theory of relativity.**

The first direct detection of gravitational waves [1], ripples in the fabric of spacetime triggered by the merger of two black holes, confirms a major prediction of the general theory of relativity (GR). This represented the culmination of a century-long speculation and decades of painstaking work, further strengthening our confidence in Einstein's theory of gravity: at least in the strong field limit, GR passes the most stringent test to date. When applied on cosmological scales, GR also successfully describes the overall dynamics of the Universe and the development of structure within it. However, as indicated by the Hubble diagram of distant Type Ia supernovae, the observed expansion history of the Universe encapsulated in the expansion rate, H(z), requires the introduction of a cosmological constant term $\Lambda$ into Einstein's equations [5]-[6]. This is a problem for quantum theory, which is not able to predict the small observationally inferred value of $\Lambda$, nor explain its stability without fine-tuning (see e.g. Ref. [4] for a review). In view of this, one may alternatively reproduce the observed H(z) by modifying the very nature of Einstein's theory.

A number of such modifications to GR have been proposed and discussed over the past decade (see Ref. [3] for a recent review). The simplest generalisation is represented by the so-called $f(R)$ models, in which the Ricci scalar curvature, $R$, in the Einstein-Hilbert action is replaced by an arbitrary function of $R$ (see Ref. [13] for a review). A virtue of $f(R)$ gravity is that it adheres to all of the pivotal principles that underpin GR, such as the equivalence principle. $f(R)$ gravity, like GR, also predicts gravitational waves that travel at the speed of light, as recently confirmed to unprecedented accuracy by the coeval detection of gravitational waves [2] (GW170817) and a γ-ray burst [15] (GRB 170817A) from the merger of two neutron stars. As such, $f(R)$ gravity is arguably the "least unnatural" and most compelling theory of gravity beyond Einstein's GR. As is the case with most viable modified gravity models, $f(R)$ gravity includes a "screening mechanism", which restores GR in dense environments, where strong constraints exist (see e.g. [16]). This means that deviations from GR, if any, can be detected only by measurements on cosmological scales, where such screening is, in general, ineffective. These are the scales probed by galaxy surveys and the statistics of large-scale structure.

A direct test of modifications to gravity on large scales is provided by measurements of the growth rate of structure, $f = d \ln D / d \ln a$, where $D(a) = \delta(a)/\delta_0$ is the linear growth factor of cosmological matter density $\rho$ perturbations $\delta = \delta\rho/\rho$ and $a$ is the scale factor of the Universe [17]. Structure growth generates galaxy peculiar velocities, i.e. motions superimposed on the pure cosmological expansion. As such, they add a Doppler contribution to the cosmological *redshift* in the observed spectrum of galaxies, which in large surveys is used to estimate galaxy distances and create 3D maps of their distribution. As a consequence, the maps and the statistics we derive from them are affected by what we call *redshift space distortions* (RSD) [7]-[8]. Containing direct information on the growth rate, this effect has been recognised in recent years as a key probe of the nature of gravity and hence of the origin of the cosmic acceleration [9].

A precise test of gravity using RSD is, however, non-trivial [11]. The growth rate is encoded in the large-scale coherent motions towards overdensities, under the hypothesis that these are still described by linear theory. However, two-point statistics, for which the RSD signal is modelled, are also affected by large random velocities inside virialized objects such as galaxy clusters and groups, which are the result of fully non-linear gravitational evolution. Such nonlinear effects can extend to fairly large scales, resulting in a complicated scale-dependent clustering signal (see [18] and references therein). Significant progress has been achieved over the past decade in modelling this (e.g. Refs. [19]-[20] and references therein), with a number of independent measurements of the growth rate of structure at different redshifts. These show general consistency with the predictions of GR[10][21]. In these works, systematic errors are, in general, carefully estimated through mock experiments; however, it remains hard to assess quantitatively to better than ~5% (in the best cases) [10] how model-dependent assumptions globally affect the results.

Here we take an alternative route, which avoids most of the current difficulties. Rather than applying non-linear corrections to our linear models of the growth of structure to match the data, we take a forward modelling approach that starts from the fully non-linear description of matter clustering provided by numerical N-body simulations of structure formation in standard and non-standard cosmologies. The key problem becomes then to associate self-consistently and unambiguously observed galaxies to dark-matter halos in the simulations. This is achieved using the so-called *sub-halo abundance matching* (SHAM) technique [22]-[23] (See **Methods** for details). Following the standard paradigm of galaxy formation, galaxies reside in DM halos (usually called *sub-halos*, as often embedded within more massive parent halos, as e.g. a group or a cluster). If a monotonic relation exists between a property of the sub-halo and a physical property of the galaxy, then there is a one-to-one match between simulated DM sub-halos and galaxies in a survey, selected according to that property. This implies, for example, that the clustering of sub-halos in a simulation can be directly and quantitatively compared to the measured clustering of a corresponding sample of galaxies. The relevant virtues of this approach are that: (a) there is no ambiguity regarding *galaxy bias*, which links the clustering (and RSD) of galaxy tracers to the underlining dark matter (DM) field; (b) non-linear clustering and motions are fully reproduced, up to the scales set by the mass resolution and volume of the simulation.

To implement SHAM in practice, the key point is to identify the specific physical property of observable galaxies that monotonically (or with very small scatter) depends on a property of the host DM halo. State-of-the-art hydrodynamic simulations like EAGLE [24]-[25], in which dissipative processes and a comprehensive physical model of galaxy formation are included, show that a tight correlation (albeit with some scatter) exists between galaxy stellar mass (the total mass of a galaxy in stars) and the peak value of the maximum circular velocity over the subhalo's merger history, $v_{peak}$. While stellar mass is derived from multi-band photometric measurements, $v_{peak}$ turns out to be a robust proxy for the maximum gravitational potential attained over the existence of a simulated sub-halo (See **Methods** for details).

We adopt the SHAM technique in this work to make a quantitative comparison of redshift-space clustering measured from the Sloan Digital Sky Survey (SDSS) main galaxy sample [14] to the predictions of two simulations, built in the standard $\Lambda$CDM scenario and in an $f(R)$ cosmology. To construct our reference data sample, we use the New York University Value-Added Galaxy Catalogue (NYU-VAGC) [26], which is an enhanced version of the SDSS Data Release 7 [14]. At a limiting magnitude $r \leq 17.60$, this catalogue is highly complete and uniform over an area of 7732 deg², including 542,432 galaxies with measured spectroscopy and a median distance of ~300 Mpc/h. From the observed *redshift* of spectral lines, *redshift space* distances are computed (note that given the low redshift of the sample, distances are only very mildly dependent on the assumed cosmology). From this catalogue we then build *volume-limited* samples that are complete in stellar-mass, following [27] (see **Methods** for details). Unlike luminosity, galaxy stellar mass is not directly obtained from the observed flux but needs to be derived from the fit of a stellar population synthesis model to multi-band photometric measurements (see. e.g. Ref. [28]). This procedure is prone to systematic errors (see e.g. Ref. [29] for a review). To mitigate these, we define our volume-limited samples in terms of a galaxy number density threshold, rather than applying a limit in stellar mass. As demonstrated in the **Methods** (and shown in Figure 2), if a sufficiently high-density sample is used, the impact of stellar mass systematic errors on clustering measurements is negligible. For this reason, our reference sample is defined by imposing a mean galaxy density $n_g = 1 \times 10^{-2} [\text{Mpc/h}]^{-3}$ (see **Extended Data Figure 1)**.

This reference data catalogue is first matched to a simulation performed in the standard GR – ΛCDM scenario. Specifically, we choose the Small MultiDark Planck simulation (SMDPL) [12], which adopts a Planck cosmology [30] with $\Omega_m = 0.3071$, $h = 0.6777$, $\sigma_8 = 0.8228$, $n_s = 0.96$. The simulation uses $3840^3$ dark matter particles in a box of 400Mpc/h side, giving a mass resolution of $9.6 \times 10^7 M_\odot/h$. Dark matter halos and subhalos in our analyses are identified using the Rockstar halo finder [31]. The high mass resolution of this simulation is crucial for the application of the subhalo abundance matching technique. A satellite subhalo close to the centre of its parent halo may lose substantial mass due to tidal stripping. At an earlier time in the simulation, a massive halo can even be completely disrupted by tidal stripping, such that at a later time it is no longer detectable. This leads to what in simulations have been called *orphan galaxies* [32], i.e. galaxies that may exist as baryonic objects today, but devoid of an identifiable DM sub-halo. Overlooking these objects in the simulation can under-estimate the predicted clustering on small scales [32]. In our case, based on Ref. [33] and given the high mass resolution of the SMDPL, our reference galaxy number density $n_g = 1 \times 10^{-2}[\text{Mpc/h}]^{-3}$ guarantees that the fraction of orphan galaxies is less than 2.6%, which is a negligible contribution. Another technical issue is that in the SHAM approch, a scatter is usually added to the relation between $v_{\text{peak}}$ and galaxy stellar mass [34]. This quantity essentially only affects the selection of subhalos near the sample mass threshold, which at the high number densities we have in our samples, has a limited impact. For this reason, in our SHAM implementation we avoid adding any scatter and, as a consequence, our model has no free parameters.

Our SHAM prediction for $f(R)$ gravity is based on a second simulation, presented in Ref. [35], coupled to an *effective halo technique* [36]-[37]. This is necessary as the circular velocity $v_{\text{cir}}$ of a baryonic particle in a subhalo in $f(R)$ gravity is not directly related to the cold dark matter mass but to an effective mass defined through a modified version of the Poisson equation [37]

$$\nabla^2 \varphi = 4\pi G a^2 \delta \rho_{\text{eff}} \;,$$

with G being Newton's constant. The simulation has a mass resolution of $1.52 \times 10^8 \text{M}_\odot/\text{h}$, i.e. comparable to that of the SMDPL simulation, which makes this the highest resolution $f(R)$ cosmological simulation to date. This is crucial for this kind of test, since the "screening mechanism", which significantly affects the velocity field in $f(R)$ cosmology, can only be accurately explored if the resolution is sufficiently high. Although the computational cost of this kind of simulation limits the box size to 64Mpc/h, which misses the long-wavelength modes (large scales) of the density field, the simulation can still produce reliable predictions for the higher-order multipoles of clustering (see **Extended Data Figure 5** and **Methods** for details), which are the ones containing most of the velocity-field information and as such the most important quantities to test gravity. In any case, we further develop a self-consistent correction for the missing large-scale modes, by building a test ΛCDM simulation with the same box size (64Mpc/h) and initial conditions as the $f(R)$ simulation. In this case, the missing fluctuations larger than the box size will be, by construction, the same for the two simulations, with the differences between their respective multipoles arguably reflecting only the different nature of gravity. The ratio of the multipoles for the two (big and small) ΛCDM simulations can then be used to re-normalise the corresponding $f(R)$ multipoles, such that they can be compared on an equal footing to the final SMDPL measurements. Details and robustness tests of this method are described in **Methods**.

From the ΛCDM simulation, we build a SHAM mock survey that fully reproduces the survey mask, geometry and wide-angle effects, volume-limited to have a number density $n_g = 1 \times 10^{-2} [\text{Mpc/h}]^{-3}$ (see **Methods**). We then compute the redshift-space two-point correlation function $\xi(r_\sigma, r_\pi)$ for the $f(R)$ and ΛCDM mock surveys and for the real data, using the standard Landy and Szalay estimator [38]; here $r_\sigma$ and $r_\pi$ give the separation of galaxy pairs split into the directions perpendicular and parallel to the line-of-sight, respectively (see **Methods).** In Figure 1 we compare the measurement from the SDSS data with that of the ΛCDM mock sample. It is remarkable how well the data and the model agree, particularly on small scales in the highly non-linear regime. This is exactly where data and models can only be compared using simulations, which is the key advantage of our approach. The corresponding two-dimensional $\xi(r_\sigma, r_\pi)$ for $f(R)$ cosmology is much noisier due to the limited box size of the simulation; to reduce the noise and get a better statistical comparison with data, we compress the information into spherical harmonics moments, namely the monopole $\xi_0$, quadrupole $\xi_2$ and hexadecapole $\xi_4$, as detailed in **Methods**. In the same section we also describe how we account for the well-known effect of the SDSS spectrograph fibre collisions on very small scales, using the so-called *truncated multipoles* [39]-[40]. The results are shown in Figure 2 for both ΛCDM and $f(R)$, compared to the SDSS measurements. As expected from Figure 1, the ΛCDM model is an excellent description also of the clustering multipoles, within $1\sigma$ statistical uncertainty of the observational measurements, given by the small error bars. The $f(R)$ prediction, however, has a significantly smaller amplitude on small scales; despite its relatively larger theoretical uncertainty, this corresponds to a more than 5-sigma discrepancy with the SDSS data on small scales ($s < 6\text{Mpc/h}$), as shown by the shaded bars.

The robustness of this result with respect to systematic errors in stellar mass estimators is also indicated by different symbols: different stellar mass models yield consistent clustering for the corresponding mass-selected samples, indicating that they do not significantly change the rank-order of the galaxies in the SHAM implementation (see also **Extended Data Figure 3**).

Overall, these results show that when data and simulations are self-consistently matched with physically motivated and robustly tested forward modelling, the ΛCDM model in the framework of general relativity provides an impressively accurate description of *both* galaxy clustering and motions on small-scales. Conversely, a representative of the family of $f(R)$ models, such as the one considered here [41] is clearly ruled out. This incarnation of $f(R)$ is characterised by a free parameter $f_{R0} = -10^{-6}$ and an index n=1. Such a small value of $|f_{R0}|$ makes this $f(R)$ model barely distinguishable from ΛCDM when using other cosmological probes, such as cluster counts or weak lensing [42]. Yet, the impact on the observed redshift-space clustering is dramatic, as we can see from Figure 2. The sensitivity of the non-linear velocity field to even tiny modifications to the laws of gravity is remarkable [43]. Our result provides a strong confirmation on cosmological scales of Einstein's general theory of relativity, significantly reducing the appeal of a modification to gravity as a solution to the conundrum of the cosmic acceleration.

[1]  Abbott, B. P. *et al*. Observation of Gravitational Waves from a Binary Black Hole Merger. *Phys. Rev. Lett.* **116**, 061102 (2016).

[2]  Abbott, B. P. *et al*. GW170817: Observation of Gravitational Waves from a Binary Neutron Star Inspiral. *Phys. Rev. Lett.* **119**, 161101 (2017).

[3]  Koyama, K. *et al*. Cosmological tests of modified gravity. *Rept. Prog. Phys.,* **79**, 046902 (2016).

[4]  Carroll, S. M. The Cosmological Constant. *Living Rev. Rel.* **4**, 1 (2001).

[5]  Perlmutter, S. J. *et al*., Discovery of a supernova explosion at half the age of the Universe. *Nature* **391**, 51 (1998).

[6]  Riess, A. G. *et al*. Observational Evidence from Supernovae for an Accelerating Universe and a Cosmological Constant. *Astron. J.* **116**, 1009 (1998).

[7]  Kaiser, N. Clustering in real space and in redshift space. *Mon. Not. R. Astron. Soc.* **227**, 1 (1987).

[8]  Peacock, J. A. et al. A measurement of the cosmological mass density from clustering in the 2dF Galaxy Redshift Survey. *Nature*, **410**, 169 (2001).

[9]  Guzzo, L. *et al*. A test of the nature of cosmic acceleration using galaxy redshift distortions. *Nature*, **451**, 541 (2008).

[10] S. Alam *et al*. The clustering of galaxies in the completed SDSS-III Baryon Oscillation Spectroscopic Survey: cosmological analysis of the DR12 galaxy sample *Mon. Not. R. Astr. Soc.* **470**, 2617 (2017).

[11] de la Torre, S. *et al*. Modelling non-linear redshift-space distortions in the galaxy clustering pattern: systematic errors on the growth rate parameter. *Mon. Not. R. Astr. Soc.* **427**, 327 (2012).

[12] Klypin, A. *et al*. MultiDark simulations: the story of dark matter halo concentrations and density profiles. *Mon. Not. R. Astr. Soc.* **457**, 4340 (2016).

[13] Felice, A. D. *et al*. $f(R)$ theories. *Living Rev. Rel.* **13**, 3 (2010).

[14] Abazajian K. N. *et al*. The seventh data release of the Sloan digital sky survey. *Astrophys. J. Suppl.* **182**, 543 (2009).

[15] Goldstein, A. *et al*. An ordinary short gamma-ray burst with extraordinary implications: Fermi-GBM detection of GRB 170817A. *Astrophys. J.* **848**, L14 (2017).

[16] Khoury, J. *et al*. Chameleon Fields: Awaiting Surprises for Tests of Gravity in Space. *Phys. Rev. Lett.* **93**, 171104 (2004).

[17] Linder, E.V. Cosmic growth history and expansion history. *Phys. Rev D* **72**, 043529 (2005).

[18] Bianchi, D. *et al*. Improving the modelling of redshift-space distortions - I. A bivariate Gaussian description for the galaxy pairwise velocity distributions. *Mon. Not. R. Astr. Soc.* **446**, 76 (2015).

[19] Scoccimarro R. *et al*. Redshift-Space Distortions, Pairwise Velocities and Nonlinearities. *Phys. Rev D* **70**, 083007 (2004).

[20] Bianchi, D. *et al*. Statistical and systematic errors in redshift-space distortion measurements from large surveys. *Mon. Not. R. Astr. Soc.* **427**, 2420 (2012).

[21] de la Torre, S. *et al*. The VIMOS Public Extragalactic Redshift Survey (VIPERS). Gravity test from the combination of redshift-space distortions and galaxy-galaxy lensing at 0.5<z<1.2. *A&A.* **608**, A44 (2017)

[22] Vale, A. *et al*. Linking halo mass to galaxy luminosity. *Mon. Not. R. Astr. Soc.* **353**, 189 (2004).

[23] Conroy, C. *et al*. Modeling luminosity-dependent galaxy clustering through cosmic time, *Astrophys. J.* **647**, 201(2006).

[24] Schaye, J. *et al*. The EAGLE project: simulating the evolution and assembly of galaxies and their environments. *Mon. Not. R. Astr. Soc.*, **446**, 521 (2015).

[25] Chaves-Montero, J. *et al*. Subhalo abundance matching and assembly bias in the EAGLE simulation. *Mon. Not. R. Astr. Soc.*, **460**, 3100 (2016).

[26] Blanton, M. R. *et al*. NYU-VAGC: a galaxy catalog based on new public surveys. *Astron. J.* **129**, 2562 (2005).

[27] van den Bosch, F. C. *et al*. The importance of satellite quenching for the build-up of the red sequence of present day galaxies. *Mon. Not. R. Astr. Soc.* **387**, 79 (2008).

[28] Bruzual, G. *et al*. Stellar population synthesis at the resolution of 2003. *Mon. Not. R.*

*Astr. Soc.*, **344**, 1000 (2003).

[29] Conroy C., Modeling the Panchromatic Spectral Energy Distributions of Galaxies. *ARA&A*, **51**, 393 (2013).

[30] Ade, P. A. R. *et al*. Planck 2015 results. XIII. Cosmological parameters. *A&A* **594**, A13 (2016)

[31] Behroozi P. S. *et al*. The ROCKSTAR Phase-space Temporal Halo Finder and the Velocity Offsets of Cluster Cores. *Astrophys. J.* **762**, 109 (2013).

[32] Moster, B. P. *et al*. Constraints on the Relationship between Stellar Mass and Halo Mass at Low and High Redshift. *Astrophys. J.* **710**, 903 (2010).

[33] McCullagh, N. *et al*. Revisiting HOD model assumptions: the impact of AGN feedback and assembly bias. *arXiv:1705.01988*.

[34] Reddick, R. M. *et al*. The connection between galaxies and dark matter structures in the local universe. *Astrophys. J.* **771**, 30 (2013).

[35] Shi, D. *et al*. Exploring the liminality: properties of halos and subhalos in borderline $f(R)$ gravity. *Mon. Not. R. Astr. Soc.* **452**, 3179 (2015).

[36] He, J.-H. *et al*. Subhalo abundance matching in $f(R)$ gravity. *Phys. Rev. Lett.* **117**, 221101 (2016).

[37] He, J.-H. *et al*. Effective Dark Matter Halo catalogue in $f(R)$ gravity. *Phys. Rev. Lett.* **115**, 071306 (2015).

[38] Landy, S.D. *et al*. Bias and variance of angular correlation functions. *Astron. J.* **412**, 64 (1993).

[39] Reid, B. A. *et al*. A 2.5% measurement of the growth rate from small-scale redshift space clustering of SDSS-III CMASS galaxies. *Mon. Not. R. Astr. Soc.* **444**, 476 (2014).

[40] Mohammad F. G. *et al*. Group–galaxy correlations in redshift space as a probe of the growth of structure. *Mon. Not. R. Astr. Soc.* **458**, 1948 (2016).

[41] Hu, W. *et al*. Models of $f(R)$ cosmic acceleration that evade solar system tests. *Phys. Rev D* **76**, 064004 (2007).

[42] Schmidt, F. *et al*. Cluster Constraints on $f(R)$ Gravity. *Phys. Rev D* **80**, 083505 (2009).

[43] Fontanot, F. *et al*. Semi-analytic galaxy formation in $f(R)$-gravity cosmologies. *Mon. Not. R. Astr. Soc.* **436**, 2672 (2013).

[44] Blanton M. R. *et al*. K-Corrections and Filter Transformations in the Ultraviolet, Optical, and Near-Infrared. *Astron. J.* **133**, 734 (2007).

[45] Yang, X. *et al*. Galaxy Groups in the SDSS DR4: I. The Catalogue and Basic Properties. *Astrophys. J.* **671**, 153 (2007).

[46] Chabrier, G. *et al*. Galactic Stellar and Substellar Initial Mass Function. *PASP*, **115**, 763 (2003).

[47] Kroupa, P. *et al*. On the variation of the initial mass function. *Mon. Not. R. Astr. Soc.*, **322**, 231 (2001).

[48] Guo, Q. *et al*. How do galaxies populate Dark Matter halos? *Mon. Not. R. Astr. Soc.* **404**, 1111 (2010).

[49] Li, C. & White, D. M. The distribution of stellar mass in the low-redshift Universe. *Mon. Not. R. Astr. Soc.* **398**, 2177 (2009).

[50] Davis, M. & Peebles, P. J. E. A survey of galaxy redshifts. V-The two-point position and velocity correlations. *Astrophys. J.* **267**, 465 (1983).

[51] Kravtsov, A. V. *et al*. The Dark Side of the Halo Occupation Distribution. *Astrophys. J.* **609**, 35 (2004).

[52] Springel, V. The cosmological simulation code GADGET-2, *Mon. Not. R. Astr. Soc.*, **364**, 1105 (2005).

[53] Teyssier, R. Cosmology Hydrodynamics with adaptive mesh refinement. A new high-resolution code called RAMSES. *Astron. Astrophys.*, **385**, 337 (2002).

[54] Knollmann, S. R. *et al*. Ahf: Amiga's Halo Finder. *Astrophys. J. Suppl.* **182**, 608 (2009).

[55] Smee, S.A. *et al*. The Multi-object, Fiber-fed Spectrographs for the Sloan Digital Sky Survey and the Baryon Oscillation Spectroscopic Survey. *Astron. J.* **146**, 40 (2013).

[56] Hawkins, E. *et al*. The 2dF Galaxy Redshift Survey: correlation functions, peculiar velocities and the matter density of the Universe. *Mon. Not. R. Astr. Soc.* **346**, 78 (2003).

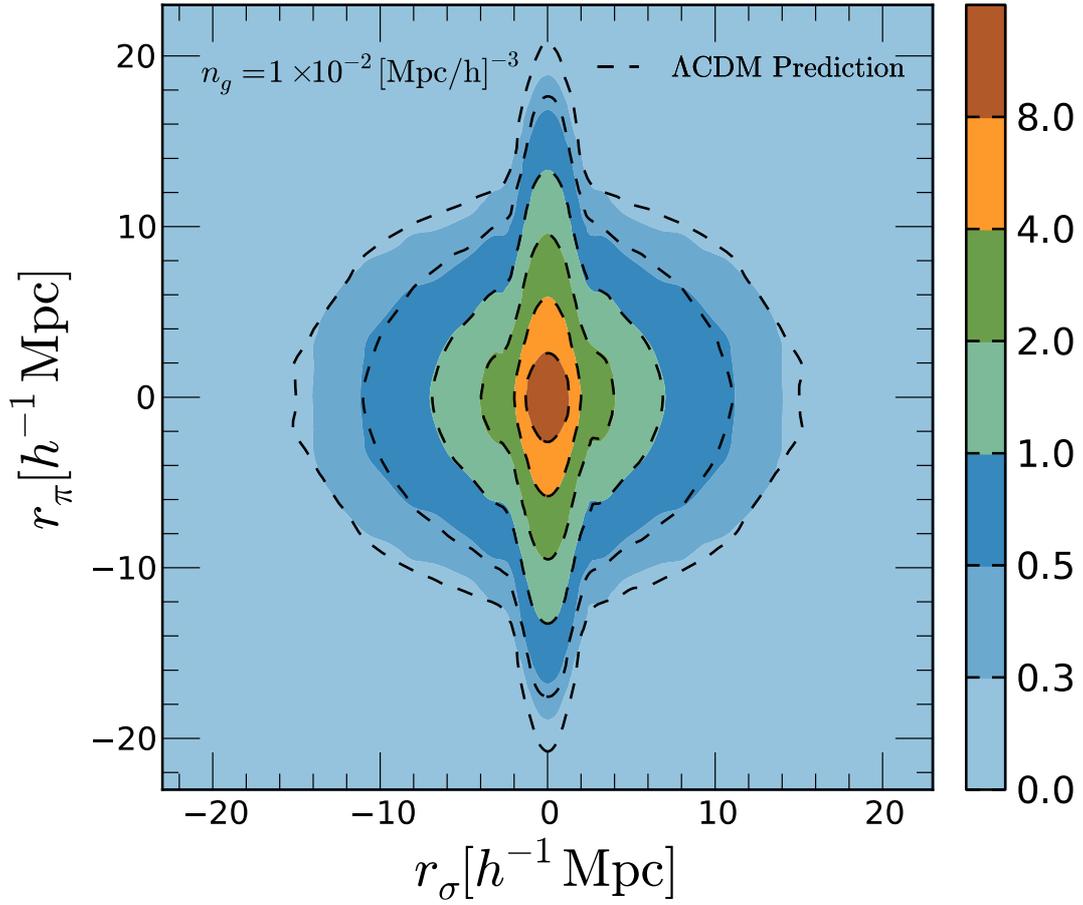

**Figure 1 | The small-scale galaxy clustering measured from the SDSS in redshift-space, compared to a realisation of the standard ΛCDM model.** The colour-coded contours show the amplitude of the two-point correlation function $\xi(r_\sigma, r_\pi)$ of a stellar-mass selected sample of galaxies from the SDSS-NYU catalogue, as a function of the transverse $r_\sigma$ and radial $r_\pi$ separation. The actual measurement for positive $r_\sigma$ and $r_\pi$ is replicated over four quadrants to highlight deviations from circular symmetry, produced by galaxy peculiar velocities that add to cosmological expansion when we use galaxy redshifts as a proxy for distance. Without peculiar motions, the contours would be perfect circles. The dashed lines give the corresponding predictions of ΛCDM, obtained from a mock galaxy survey fully mimicking the real data, based on the SMDPL N-body simulation. Dark-matter halos in the simulation and galaxies have been matched through an implementation of Sub-Halo Abundance Matching (SHAM) without free parameters (no scatter). The ΛCDM predictions agree impressively well with the observations in redshift space, especially for the so-called "Fingers-of-God" feature at

small $r_\sigma$, i.e. the stretching of the contours along the line of sight produced by high-velocity galaxies in groups and clusters. This highlights the advantage of our forward modelling approach using simulations to allow a comparison of galaxy clustering and motions for data and models into the fully non-linear regime.

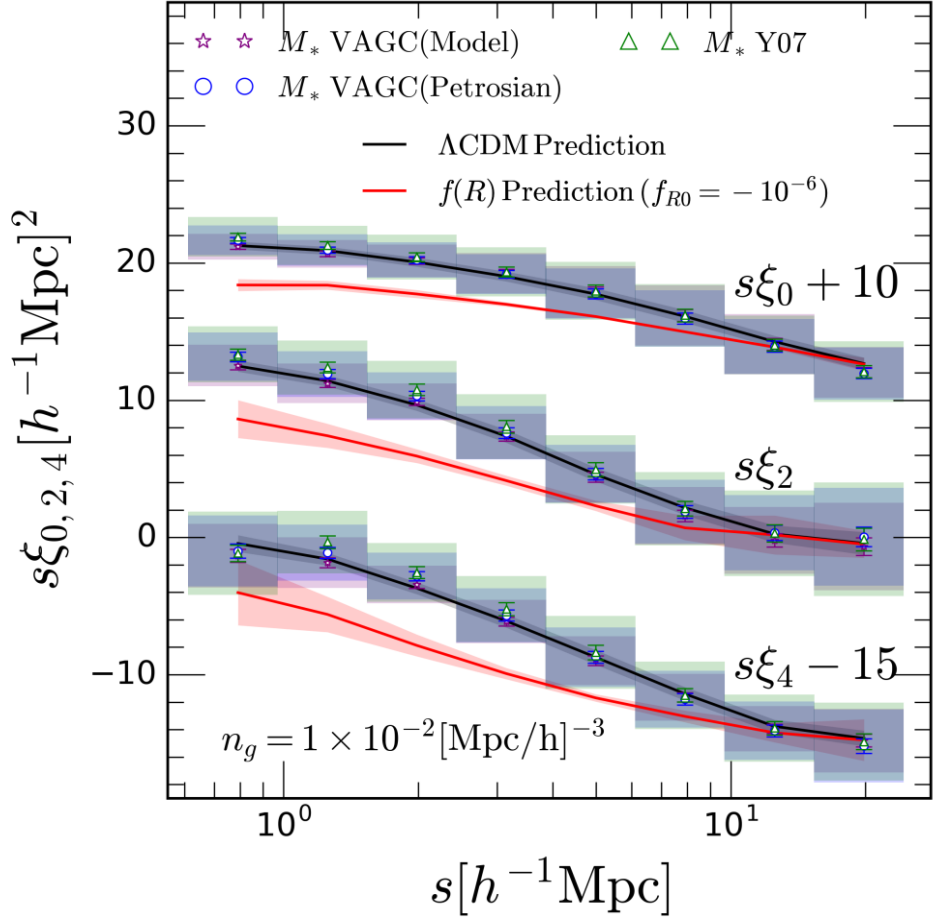

**Figure 2 | Multipoles of the redshift space two-point correlation $\xi(r_\sigma, r_\pi)$.** Galaxy clustering predicted by standard ΛCDM (black solid line) and the $f(R)$ model (red solid line) are compared with the SDSS measurements (symbols with errors), in terms of the monopole $\xi_0$, quadrupole $\xi_2$ and hexadecapole $\xi_4$ (multiplied by the *redshift space* separation s for convenience). These were estimated using the "truncation" technique (see **Methods**). As suggested by Figure 1, ΛCDM is in excellent agreement with observational measurements, while the $f(R)$ model is not. To test the robustness of this result with respect to different estimates of galaxy stellar mass, we compare three methods: a template-fit method [44] as adopted in the NYU catalogue with the SDSS model magnitudes (stars), the same, but using SDSS Petrosian magnitudes (circles), and a single-color method [45] (triangles). Clearly, changing the estimates of stellar mass does not change the results appreciably; this happens because different mass estimations approximately preserve the same rank-order of galaxies. Small error bars show the 1σ statistical error estimated using jack-knife re-sampling with 133

realizations, while the large shaded bars indicate a 5σ statistical error. The black and red shaded corridors indicate the 1σ uncertainty in the theoretical predictions of the two cosmological models.

**Correspondence** Correspondence and requests for materials should be addressed to Jian-hua He (jianhua.he@durham.ac.uk)

**Competing Interests** The authors declare that they have no competing financial interests

**Author Contributions** All authors contributed to the development and writing of this paper. J.H. led the data analysis. B.L. conducted and provided the simulations. L.G. led the writing of the paper. C.M.B. and J.H. conceived the idea of data analysis.

**Acknowledgements** JH is supported by a Durham co-fund Junior Research Fellowship; JH and BL acknowledge support by the European Research Council (ERC-StG-716532-PUNCA); LG acknowledges support by the European Research Council (ERC-AdG-291521-Darklight) and by the Italian Space Agency (ASI Grant I/023/12/0); BL and CMB are also supported by UK STFC Consolidated Grants ST/P000541/1 and ST/L00075X/1.

**Data and Code Availability**. The observational datasets and the SMDPL simulation used in this study are available from the publicly accessible NYU-VAGC catalogue and the MultiDark Database. The halo catalogue of the f(R) simulation and the codes that support the findings of this study are available from the corresponding author upon reasonable request.

## Methods

**Observational data.** Our analysis uses the New York University Value-Added Galaxy Catalogue (NYU-VAGC) [26], which is an enhanced version of the Sloan Digital Sky Survey (SDSS) main galaxy sample Data Release 7 [14]. Specifically, we use the **NYU-VAGC bbright** catalogue, which has a fairly homogenous *r*-band Petrosian magnitude limit of $r \leq 17.60$ over the whole survey footprint. The catalogue covers an area of $7732 \text{ deg}^2$; of these, $144 \text{ deg}^2$ are masked due to bright stars. Galaxies in this catalogue are mainly located in a contiguous region in the north Galactic cap. We also include the three strips in the south Galactic cap, which account for about 10% of the total number of galaxies, for a total of $N_t = 594,307$ photometrically selected galaxies. Of these, $N_s = 542,432$ have a reliable spectroscopic redshift. All samples analysed here are limited to z > 0.02.

**Volume-limited samples complete in stellar mass.** We follow the method proposed in Ref. [27] to construct volume-limited samples that are complete in stellar-mass. Since blue galaxies have a lower stellar mass-to-light ratio $M_*/L$, for a given stellar mass, a blue galaxy can be detected to a higher redshift than a red one (see Extended Data Figure 1). This implies that the flux limit of a flux-limited survey translates into a stellar mass limit $M_{*\min}(z)$ which is higher for red galaxies than for blue ones. As shown in the same figure, below that limit red galaxies disappear, and a sample naively selected in stellar mass would be biased towards blue objects. Clearly, if we choose a minimum mass limit defined as that corresponding to the detected object in the sample that has the most extreme red colour, all other (bluer) galaxies of the same stellar mass will be more luminous and so will also be in the sample. As such, the resulting sample will be statistically complete in stellar mass. In practice, the maximum stellar mass-to-light ratio in the sample (corresponding, for what we just said, to the reddest objects) can be estimated from the data [27]. Extended Data Figure 2 shows the *r*-band stellar mass-to-light ratio of galaxies in the NYU catalogue for different stellar mass models (see next section). The solid lines are our estimates of the maximum stellar mass-to-light ratio as a function of the absolute *r*-band magnitude $M_r$, namely $M_{*\min}(M_r)$. The stellar mass limit, above which the sample is complete in stellar mass, is then given by $M_{*min}(M_{rmax}(z))$, where $M_{rmax}(z)$ is the maximum (i.e. faintest) *r*-band absolute magnitude $M_r$ that can be seen at redshift z given an apparent flux limit $m_r$. The solid curves in Extended Data Figure 1 show the mass limit $M_{*min}(z)$ obtained in this way for different stellar mass models. The corresponding horizontal and vertical lines show stellar mass and *redshift* limits ($z_{max}$) for 6 volume-limited sub-samples that are complete in stellar-mass.

**Stellar mass estimate systematic uncertainties.** Unlike luminosity, a galaxy's stellar mass cannot be directly measured but has to be derived from a fit to its Spectral Energy Distribution (SED), using a stellar population synthesis model (e.g. Ref. [28]), a modelling process which is prone to systematic errors. There are two main sources of errors. One lies in the theoretical uncertainty, in particular in the choice of the stellar initial mass function (IMF) (see Ref. [29] for a review). Choosing a Chabrier [46] or Kroupa IMF [47], has a significant impact on the amplitude of the stellar mass function. The second major source of systematic uncertainty lies in the way the total flux of a galaxy is estimated from the imaging data. In SDSS, the aperture used to estimate the flux in all five photometric bands (*u, g, r, i, z*) is set by the galaxy surface brightness profile as measured in the *r*-band alone. This defines an *r*-band Petrosian radius $r_p$. The total flux in all bands is then obtained by integrating out to twice this value, $2r_p$. This aperture is large enough to contain virtually all flux for objects with an exponential surface brightness profile. For objects with a de Vaucouleurs profile, however, this only typically includes 80% of the flux. As such, a significant fraction of light can be lost and SDSS Petrosian magnitudes tend to underestimate the total flux of galaxies with such a profile, a problem which is particularly acute for massive elliptical galaxies (see e.g. Ref. [48]).

To mitigate these systematics, we define our galaxy samples in terms of their number densities, rather than through thresholds in stellar mass. The idea is to keep the rank-order of galaxies stable. Changes in the IMF shift the absolute value of a galaxy stellar mass, while not significantly changing its relative rank-order. Therefore, by selecting galaxies in terms of number densities, the choice of IMF has little impact on the selected samples and, most importantly, does not perturb the rank-order in the sub-halo abundance matching.

In addition, we use galaxy samples with a high mean number density, which has a valuable effect of reducing the impact of uncertainty (scatter) in the estimated stellar mass between different models. Although, as shown in Ref. [49], if the same stellar IMF is used, the overall distribution of the estimated stellar masses in the sample remains stable when changing the model, on a galaxy-by-galaxy basis there is still significant scatter. This scatter, however, only affects the sample definition near the mass threshold chosen. If the sample is very sparse, the fraction of objects going in and out of the sample due to scatter can be significant. However, if the mean number density of the selected samples is sufficiently high, this fraction will be small compared to the bulk of the sample.

We test directly the effectiveness of such a strategy by applying three different stellar mass models to our galaxy samples. The first one is the default model used in the NYU catalogue, which is based on the five-band SDSS Petrosian photometry and a fit to templates from a stellar population synthesis model [44]. The template fit yields the galaxy stellar mass-to-light ratio $M_*/L$, and the stellar mass can then be obtained by multiplying by its luminosity. In order to address the impact of the fixed $r$-band aperture on the estimated stellar mass, we adopt the same template fitting method, but use the SDSS model magnitudes, instead of the Petrosian ones. The former are obtained by fitting the (Point-Spread-Function convolved) exponential and de Vaucouleurs profiles to a galaxy and then adopting the one giving the best $\chi^2$. In contrast to the SDSS Petrosian magnitudes, SDSS model magnitudes better account for the loss of flux due to the fixed aperture. Finally, we also consider a third model based on a single-colour method, following [45]. In this case, the stellar mass is given by

$$\log_{10}(M_*/[h^{-2}M_\odot]) = -0.406 + 1.097[^{0.0}(g-r)] - 0.4(^{0.0}M_r - 5\log_{10} h - 4.64),$$

where $^{0.0}M_r - 5\log_{10} h$ is the absolute magnitude that is K-corrected and evolution corrected to redshift zero, as

$$^{0.0}M_r - 5\log_{10} h = m_r - DM(z) - k_{0.0}(z) + 1.62z.$$

Here $k_{0.0}(z)$ is the K-correction to redshift zero and $DM(z)$ is the distance modulus

$$\mathrm{DM}(z) \equiv 5\log_{10}(D_L/[\mathrm{Mpc}/h]) + 25,$$

with $D_L$ being the luminosity distance in Mpc/h. Note that in this estimator, the stellar mass is only explicitly dependent on the rest-frame $^{0.0}(g-r)$ colour and magnitude. It is also important to note that this single-colour estimator implicitly assumes the Kroupa IMF [47], while the default NYU catalogue uses the Chabrier IMF [46]. Extended Data Figure 2 compares the stellar mass-to-light ratio obtained for the three different models. The right panel shows the significant differences between the single-colour estimator and the photometric template-fit method: estimated stellar masses differ not only on a galaxy-by-galaxy basis, but also in the resulting global $M_*/L$ relation.

**Two-point correlation function estimators.** To estimate the redshift-space two point correlation function $\xi(r_\sigma, r_\pi)$ [50] (where $r_\sigma$ and $r_\pi$ are the separations of galaxy pairs perpendicular and parallel to the line-of-sight direction, respectively), we use the well-known Landy and Szalay estimator [38]. To reveal deviations from isotropy, $\xi(r_\sigma, r_\pi)$ can be conveniently expanded in terms of Legendre polynomials, as

$$\xi_l(s) = \frac{2l+1}{2}\int_{-1}^{1} d\mu\, \xi(s,\mu) P_l(\mu),$$

where $P_l(\mu)$ is the Legendre polynomial of order l, $s = \sqrt{r_\sigma^2 + r_\pi^2}$ and $\mu = r_\pi/s$. In Extended Data Figure 3 we plot the monopole $\xi_0$, quadrupole $\xi_2$ and hexadecapole $\xi_4$ of $\xi(r_\sigma, r_\pi)$ measured from different galaxy samples with varying mean densities (different colours) and based upon different stellar mass models (different line styles). For samples with very low mean densities, such as $n_g = 5 \times 10^{-4}[\mathrm{Mpc}/h]^{-3}$ and $n_g = 1 \times 10^{-3}[\mathrm{Mpc}/h]^{-3}$ (cyan and magenta groups of lines), the systematic errors in stellar mass estimates do affect the resulting galaxy clustering. This is precisely the effect due to a significant fraction of objects in the sample, with stellar mass close to the mass selection threshold, that can move in or out of the sample depending on the method used to estimate their masses. However, this effect is minimised for much denser samples, corresponding to $n_g = 5 \times 10^{-3}[\mathrm{Mpc}/h]^{-3}$ and $n_g = 1 \times 10^{-2}[\mathrm{Mpc}/h]^{-3}$ (yellow and olive), with the clustering properties remaining substantially unchanged with respect to the mass estimator used.

In addition, in SHAM (see next section), with a higher mean density we move further down the galaxy stellar mass function, thus increasing the fraction of satellite galaxies, i.e. the population dominating non-linear RSD effects. This population of high-speed satellites is also expected to be free of the so-called *velocity bias*, which can be a potential issue for central galaxies [39]. For this reason, for our reference sample we choose $n_g = 1 \times 10^{-2} [\text{Mpc/h}]^{-3}$, which is dense and also has a large enough volume to allow a robust clustering measurement.

**Sub-halo abundance matching.** The SHAM method is based on the simple assumption that there is a monotonic relationship between a property of dark matter sub-halos and a property of the corresponding galaxies. Usually, these are identified respectively with the subhalo circular velocity (as a proxy for the self-gravity of the halo) and the galaxy total stellar mass. In its original form [51], the actual maximum circular velocity of subhalos measured in the simulation (hereafter $v_{\text{max}}$) was used. In this case, however, it has been shown that the corresponding SHAM catalogue of galaxies under-predicts the observed clustering on small scales [51]. The reason is that, in contrast to its more tightly bound stellar component, the dark matter halo of a galaxy can easily be disrupted by the tidal field of a nearby or parent massive halo. A better proxy is shown to be provided by the subhalo's *maximum circular velocity at the epoch of accretion* (hereafter $v_{\text{acc}}$), before this disruption happens [23]. This allows us to recover, e.g., galaxies associated with sub-halos in the central region of a host halo. Even better results are obtained if one uses the *peak value of the maximum circular velocity over the subhalo's merger history* [34] (hereafter $v_{\text{peak}}$). The reason for this is that at the epoch of $v_{\text{peak}}$, the subhalo has the strongest binding force and, hence, is most stable against tidal stripping. One thus expects its properties to be more tightly correlated with the galaxy stellar mass at this epoch. However, as shown in [25], during the lifetime of a subhalo, $v_{\text{peak}}$ could sometimes show spikes, which might not reflect its typical status during most of its existence. This analysis, based on the state-of-the-art hydrodynamical simulation EAGLE [24], shows that the strongest correlation with galaxy stellar mass is obtained by using the highest value of the circular velocity satisfying a relaxation criterion (hereafter $v_{\text{relax}}$). At the same time, however, $v_{\text{relax}}$ is shown to be only marginally better than $v_{\text{peak}}$ in reproducing the simulated galaxy clustering. Furthermore, baryonic effects are shown to introduce only a small perturbation in a $v_{\text{peak}}$ ranked halo catalogue and have a limited impact on the positions of dark matter halos. This results in an overall accuracy in the predicted redshift-space galaxy clustering that is better than 10% above $1\text{Mpc/h}$, when using a $v_{\text{peak}}$ ranked halo catalogue in building the SHAM [25].

**Numerical stability of SHAM predictions.** The SHAM method is based on sub-halos and their merger histories, which are in turn derived from high-resolution N-body simulations. It is, therefore, important to test the robustness of SHAM predictions against different N-body codes, halo finders and methods of constructing halo merger trees. In order to do this, we perform a test simulation with $1024^3$ dark matter particles in a box of 150 Mpc/h side. In our test simulation, rather than using the GADGET code [52] as adopted in the SMDPL simulation, we instead use RAMSES [53]. RAMSES uses a multigrid relaxation method for solving the Poisson equation, which is different from the Tree-PM method used in GADGET. Moreover, we use the *Amiga Halo Finder* (AHF) [54] to identify halos and construct the halo merger tree using the MERGERTREE code which is a part of the AHF package. Extended Data Figure 4 compares the multipoles of the two-point correlation function of our test simulation (dashed lines) to those obtained from the SMDPL simulation (solid lines). In both cases, the multipoles are estimated using the *distant observer approximation* (see **Survey geometry and wide-angle effects**). Despite the significant differences in the numerical methods used, the two simulations yield very similar predictions for the clustering of derived SHAM galaxies, when a large enough number density, $n_g = 1 \times 10^{-2} [\text{Mpc/h}]^{-3}$, is adopted, as in the main paper.

**SHAM predictions in f(R) gravity.** Unlike in the $\Lambda$CDM case, the circular velocity $v_{\text{cir}}$ of a baryonic particle in a sub-halo in $f(R)$ gravity is not directly related to the true cold dark matter mass of the sub-halo but to an effective mass which is defined through a modified version of the Poisson equation

$$\nabla^2 \varphi = 4\pi G a^2 \delta \rho_{\text{eff}}$$

with G being Newton's constant [37]. The effective energy density $\rho_{\text{eff}}$, by definition,

incorporates all the effects of modified gravity. The circular velocity is then given by $v_{\rm cir}^2(r) = \frac{GM_{\rm eff}(<r)}{r}$ where $M_{\rm eff}(<r)$ is the effective mass enclosed within radius $r$ for a dark matter halo. In practice, it is more convenient to calculate the circular velocity for each dark matter halo in an $f(R)$ simulation using the effective halo catalogue technique, described in Ref. [36]-[37].

The $f(R)$ simulation used in our analyses is the one described in Ref. [35]. The simulation has $512^3$ dark matter particles within a box of 64Mpc/h a side; this allows us to reach a mass resolution of $1.52 \times 10^8 M_\odot/h$, which represents the highest resolution cosmological simulation to date for such an $f(R)$ model [41]. However, due to its limited box size, the predicted galaxy clustering cannot be directly compared to observations, since missing long-wavelength modes on scales larger than the box size also have an effect on small-scale clustering. To overcome this, we ran a further ΛCDM simulation with the same box size and – most importantly – the same initial conditions as the $f(R)$ simulation. This allows a comparison of the small-scale behaviour of the two models on equal footing: the missing long-wavelength Fourier components on scales larger than the box size will be, by construction, the same for the two simulations. As such, we expect the ratio of the two-point correlation function multipoles (monopole $\xi_0$, quadrupole $\xi_2$ and hexadecapole $\xi_4$) of the SHAM galaxy catalogues built from these two simulations to be virtually independent of the box size and reflect only the differences between the intrinsic gravity models. Thus, in practice, the full $f(R)$ predictions to be compared to the full SMDPL (Figure 2) are obtained as

$$\left(\xi_{0,2,4}^{f(R)}\right)_{\rm True} = \left(\frac{\xi_{0,2,4}^{f(R)}}{\xi_{0,2,4}^{\Lambda CDM}}\right)_{64{\rm Mpc/h}} \xi_{0,2,4}^{\rm SMDPL} = \left(\xi_{0,2,4}^{f(R)}\right)_{64{\rm Mpc/h}} \frac{\xi_{0,2,4}^{\rm SMDPL}}{\left(\xi_{0,2,4}^{\Lambda CDM}\right)_{64{\rm Mpc/h}}},$$

where $\xi_{0,2,4}^{f(R)}$ and $\xi_{0,2,4}^{\Lambda CDM}$ are obtained from the 64Mpc/h box simulations as shown in Extended Data Figure 5.

From this figure we can see that the box size affects mainly the monopole $\xi_0$, while the quadrupole $\xi_2$ and hexadecapole $\xi_4$ are essentially preserved. This is very important, beyond any correction we may apply, as the latter are the quantities that specifically measure the deviation from an isotropic distribution produced in redshift-space by the peculiar velocities of galaxies (if there were no peculiar velocities, there would be no RSD effects and only the monopole would be non-zero). As such, higher order multipoles are, in the first place, less sensitive to the underlying real-space positions of galaxies. This is particularly true on small scales, where the RSD effects are dominated by random motions of high speed galaxies, in contrast with the coherent motion of galaxies on large scales. They thus contain most of the velocity-field information and are the most robust quantities to test gravity. In view of this, we further remark that in this kind of comparison the leading consideration for an $f(R)$ simulation is its mass resolution, rather than the box size: the strength of gravity and the velocity field are significantly affected by the $f(R)$ "screening mechanism", which can only be accurately explored if the mass resolution is sufficiently high, as in our simulation.

The "screening mechanism" plays an important role in SHAM predictions for $f(R)$ gravity. Very massive main (distinct) halos in $f(R)$ gravity are usually screened. Subhalos in these main halos would feel the same strength of gravity as in standard gravity. However, less massive main (distinct) halos are usually unscreened, so that subhalos with similar (or even slightly smaller) true dark matter mass can have higher circular velocities due to enhanced gravity. As in SHAM we select subhalos using circular velocity (which is related to the effective density field and is tightly correlated to a galaxy's stellar mass in the processes of galaxy formation), the subhalos in the less massive unscreened main (distinct) halos will be selected. *For a fixed number density of halos*, this leads to the relatively weaker clustering of SHAM predictions in $f(R)$ gravity, which is in contrast with the ΛCDM case (see Figure 2 in the main text and also see Ref. [36]).

**Survey geometry and wide-angle effects.** Given the significantly different box sizes between the SMDPL and our $f(R)$ simulations, implementing redshift-space effects on their corresponding SHAM catalogues of artificial galaxies requires two different approaches. Specifically, in the case of the SMDPL, given its large volume, we can build realistic SHAM mock surveys that fully reproduce the geometry and angular mask of the real SDSS data. But for the $f(R)$ simulation, due to its limited box size, a more idealised approximation has necessarily to be used. In this section we test the robustness of the approach we used to implement redshift-space effects in the $f(R)$ simulation and clearly identify the range of scales where the $f(R)$ measurements can be compared on an equal footing to the LCDM simulation predictions and the real data.

Our "standard cosmology" $\Lambda$CDM simulated galaxy sample is built from the very large box of the SMDPL simulation [12]. In this case, we can construct a full-sky SHAM mock SDSS survey by collating 8 replicas of the box and placing the observer at the centre of this super-box. Redshift distortion effects are then obtained for each SHAM galaxy by projecting its velocity along the line-of-sight to the observer. Note that the large box size of the SMDPL simulation would be by itself sufficient for our purposes, given that we are only interested in small scales (< 20 Mpc/h). However, combining 8 replicas we can more easily accommodate the irregular geometry of the real SDSS data. This is an important check, as our SDSS volume-limited samples are in general fairly shallow and characterised by an irregular geometry. Their median redshift is only around z~0.1, i.e. ~250Mpc/h in a standard $\Lambda$CDM cosmology with $\Omega_m = 0.3$, with a large fraction of the galaxies closer than this distance. From the SMDPL simulation, we can thus build fully realistic *mock surveys* that reproduce the SDSS data selection function without introducing simplifying assumptions.

As anticipated, such an approach is not possible for an $f(R)$ cosmology. Compared to $\Lambda$CDM ones, $f(R)$ simulations are rather expensive, typically requiring 20 times more CPU time than a $\Lambda$CDM simulation with the same box size and mass resolution. This severely limits the maximum box achievable for an $f(R)$ simulation with sufficiently high mass resolution, given the current state-of-the-art in super-computers. As such, it is currently not possible to build an $f(R)$ SHAM mock sample that is large enough to accommodate the volume of a survey like the SDSS we are using here. Given this situation, to analyse RSD effects in an $f(R)$ cosmology, we adopt the commonly used approach known as the *distant observer approximation*. This is the assumption originally adopted to derive the classic linear model of RSD [7], which assumes that the box is at such a large distance, compared to its size, that both sides of the box along one direction can be considered parallel to the line-of-sight. This approach was used to produce the results of Extended Figures 4 and 5, as explicitly indicated by the label. These figures are fully self-consistent as we are comparing quantities from simulated samples treated in the same way (e.g. placed at the same, large distance). We test the robustness of this approximation against the geometry of our specific SDSS samples by applying both approaches to the SMDPL simulation. In this case, in fact, we can both build a full *mock survey* and use the distant observer approximation. The comparison of the two outcomes is shown in Extended Data Figure 6. Again, there are some noticeable effects on the monopole $\xi_0$; however, over the range 2~10 Mpc/h both the quadrupole and the hexadecapole (i.e. the two most important statistics in our gravity test), show little difference between the two ways of implementing the RSD effects, indicating that our comparisons of RSD effects in the $f(R)$ and $\Lambda$CDM SHAM catalogues with the corresponding SDSS data are robust.

**Fibre collisions mitigation.** In the SDSS twin multi-object fibre spectrographs on the 2.5 m aperture Sloan Telescope at Apache Point Observatory, two fibres cannot be placed closer than 55″ on the spectroscopic plate in the same observation, due to the physical size of the fibre plugger [55]. Thus, any two galaxies separated by 55″ or less cannot be observed simultaneously. In the SDSS observing strategy, this effect is alleviated by increasing the overlaps of adjacent tiles, such that close pairs can be targeted from different observations in the overlaps. However, there are still about 7% of the galaxies in close angular pairs that do not have a measured redshift in the survey. These missing fibre-collided pairs of galaxies introduce a systematic effect on two-point statistics, which cannot be simply accounted for through a homogeneous weighting, since the missing objects are not randomly distributed. One way to correct for this is to use an angular weight based on the ratio, as a function of separation, of the numbers of observed pairs to the total number of targets in the original survey parent catalogue [56]. This method works for a flux-limited sample, but cannot be directly generalized to sub-samples, such as volume-limited ones, at least without any assumptions [39]. A better way, which we adopt here, is – rather than using the conventional multipole expansion – to adopt the so-called *truncated multipoles* as proposed in [39] and [40]. These are defined as

$$\xi_l(s) = \frac{2l+1}{2} \int_{-1}^{1} d\mu W(s,\mu)\xi(s,\mu)P_l(\mu),$$

where $W(s,\mu)$ is a mask used to exclude the unreliable small-scale measurements from the integration

$$W(s,\mu) = \begin{cases} 1, & r_\sigma \geq 0.2 \text{Mpc/h} \\ 0, & r_\sigma < 0.2 \text{Mpc/h} \end{cases}.$$

The choice of the truncation scale of $0.2\text{Mpc/h}$ is very conservative, since at the maximum redshift of our galaxy data ($z \sim 0.1$ for the $n_g = 1 \times 10^{-2}[\text{Mpc/h}]^{-3}$ sample), the fibre angular scale of $55''$ corresponds to $r_\sigma \sim 0.08\text{Mpc/h}$. Note also that the mask $W(s,\mu)$ is used only to obtain the multipoles of $\xi(s,\mu)$. The two-dimensional correlation function $\xi(s,\mu)$, in the first place, is calculated in the usual way, using galaxy pairs from all scales. This is important, because the missing galaxies in close pairs do impact clustering on large scales as well. However, the missing power on large scales can be corrected for by properly down-weighting randoms, as implemented in our estimate of $\xi(s,\mu)$.

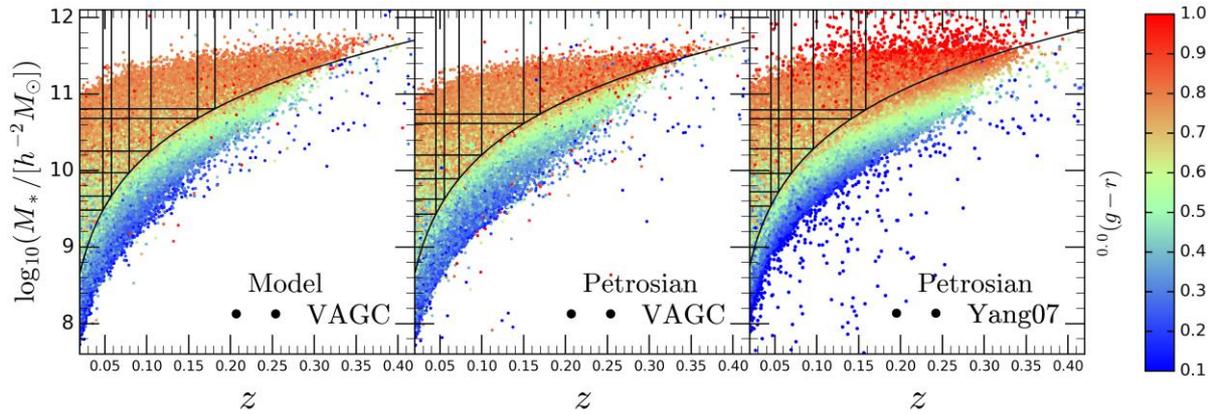

**Extended Data Figure 1.** | **The distribution of stellar masses as a function of redshift for the SDSS sample, comparing three different estimators**. The colour of each point corresponds to the galaxy rest-frame $^{0.0}(g-r)$ colour (given by the side bar); the labels indicate the different methods used to compute stellar masses in each of the three panels, as described in the text (**Methods**). Vertical and horizontal solid lines give the redshift and stellar mass limits for the volume-limited samples explored in our work. In all cases, they are limited by the solid curved line, which describes the lower completeness limit in stellar mass that has been estimated for each redshift. As discussed in the text, red objects start disappearing first when moving below this limit.

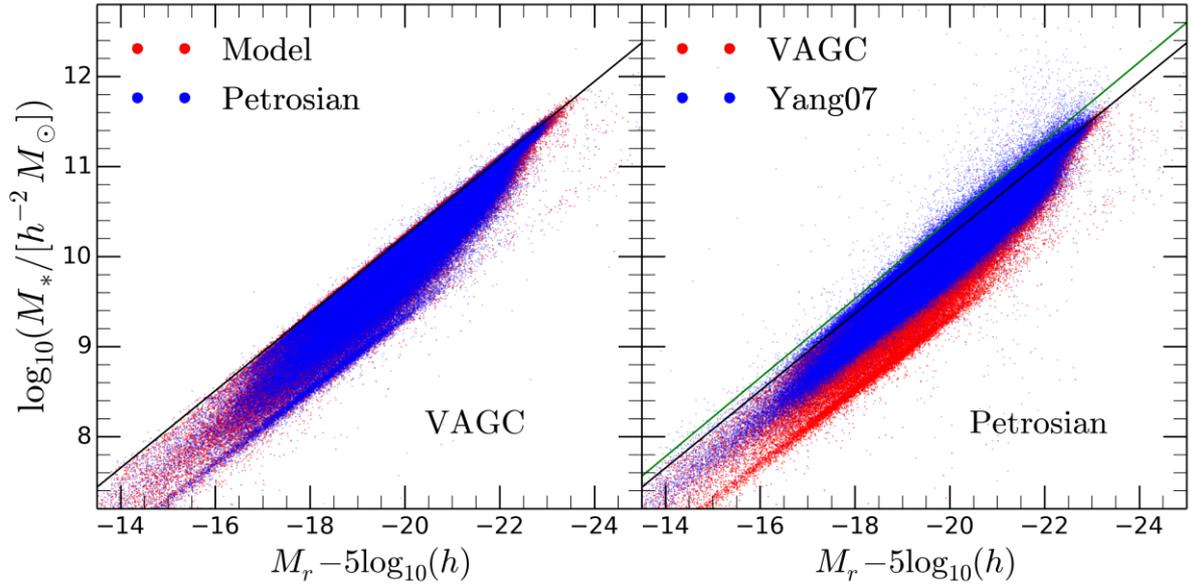

**Extended Data Figure 2. | The relationship between stellar mass $M_*$ and luminosity in the $r$-band ($M_r$) for our SDSS galaxy sample, varying the stellar mass estimator.** *Left*: results when using Petrosian and SDSS Model magnitudes. Due to the fixed $r$-band aperture, Petrosian magnitudes tend to underestimate the total flux for galaxies characterised by a de Vaucouleurs surface brightness profile (i.e. early types). SDSS Model magnitudes partly account for these differences. The two types of magnitude yield very similar results for low stellar mass galaxies, but noticeable differences emerge for massive galaxies. *Right*: same plot now comparing the single-colour method (Yang07) [45] and the template-fit method. Here differences appear not only on an object-by-object basis, but also in the overall distribution: overall, stellar mass estimates using the Yang07 model are much higher than in the default NYU catalogue. Note that the former model assumes a Kroupa IMF, while the NYU catalogue adopts a Chabrier IMF.

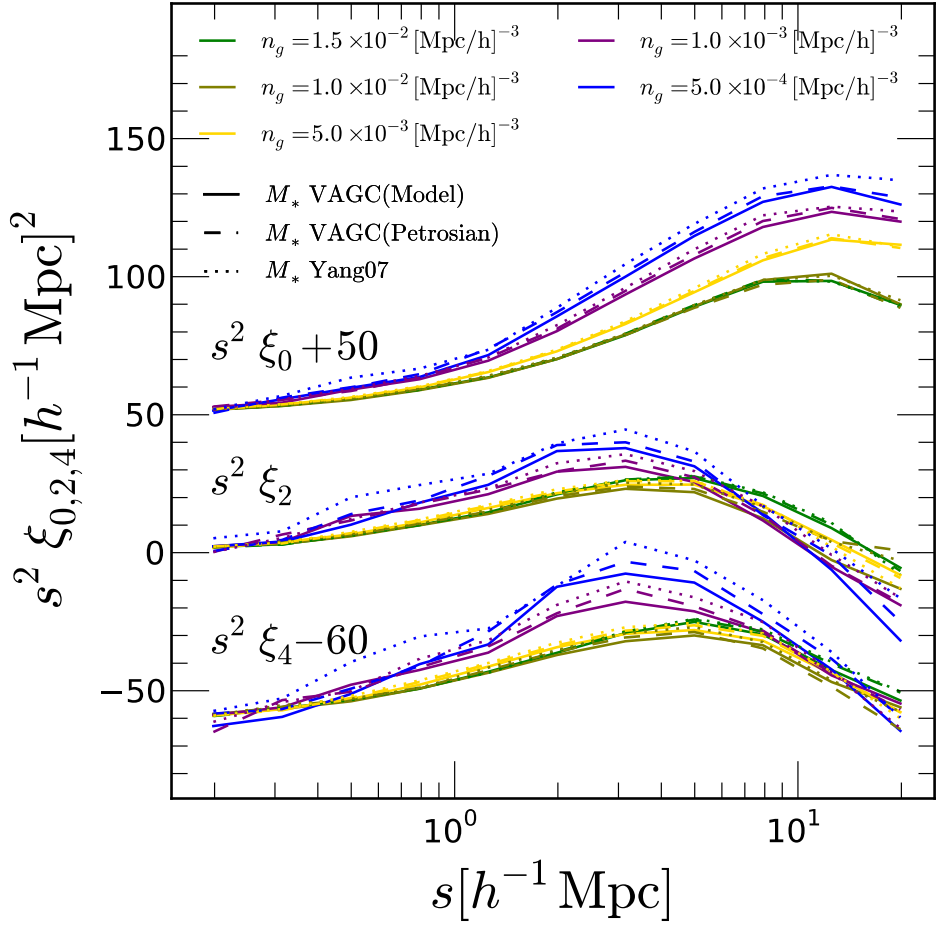

**Extended Data Figure 3. | The impact of stellar mass systematic errors on the clustering of mass-selected samples.** From top to bottom, the three groups of lines give the measured (un-truncated) monopole $\xi_0$, quadrupole $\xi_2$ and hexadecapole $\xi_4$ of the redshift space two-point correlation function $\xi(r_\sigma, r_\pi)$, for galaxy samples selected by imposing different mean number densities (corresponding essentially to different mass thresholds, as described in the text), described by different colours. Different line styles show the impact of using different stellar mass models to obtain stellar masses. For higher-mass, sparser samples (i.e. low number densities such as $n_g = 5 \times 10^{-4} [\text{Mpc/h}]^{-3}$ and $n_g = 1 \times 10^{-3} [\text{Mpc/h}]^{-3}$, blue and magenta lines) this has a noticeable effect on the measured clustering. However, for denser samples ($n_g = 5 \times 10^{-3} [\text{Mpc/h}]^{-3}$ and $n_g = 1 \times 10^{-2} [\text{Mpc/h}]^{-3}$, yellow and olive lines), the effect is much less significant and the multipoles in redshift space are virtually independent of the mass model adopted, which is what we want to obtain a robust test of the gravity models. Our cosmological analysis is thus performed by matching the halos in the simulation to the SDSS mass-selected sample with $n_g = 1 \times 10^{-2} [\text{Mpc/h}]^{-3}$.

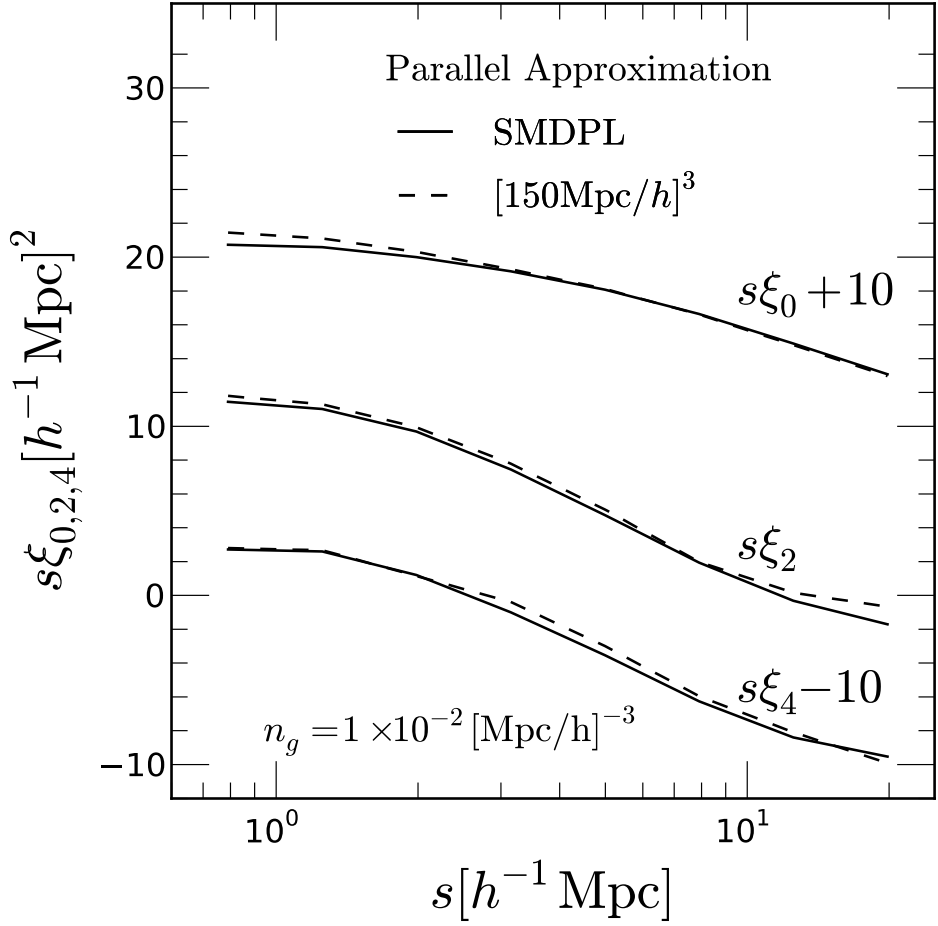

**Extended Data Figure 4. | The numerical convergence of SHAM predictions for ΛCDM.** We demonstrate here the robustness of our clustering predictions against changes in both the N-body code and the halo finder. The solid lines give the two-point correlation function multipoles for our reference SHAM sample with $n_g = 1 \times 10^{-2} [\text{Mpc}/h]^{-3}$ built from the SMDPL simulation; the dashed lines give the same statistics for a sample built from an independent simulation with the same cosmology. This test simulation has been run using the RAMSES code [53] and includes $1024^3$ dark matter particles in a box of 150 Mpc/h along one side. Sub-halos are then identified using a different code, namely the *Amiga Halo Finder* (AHF) [54] and merger trees are constructed using the MERGERTREE code which is a part of the AHF package (see text in **Methods**). Despite the significant differences in the numerical methods used, this test shows that for a sample with high number density, such as the one we use in the main paper, the two simulations yield very similar SHAM predictions for the clustering in redshift-space.

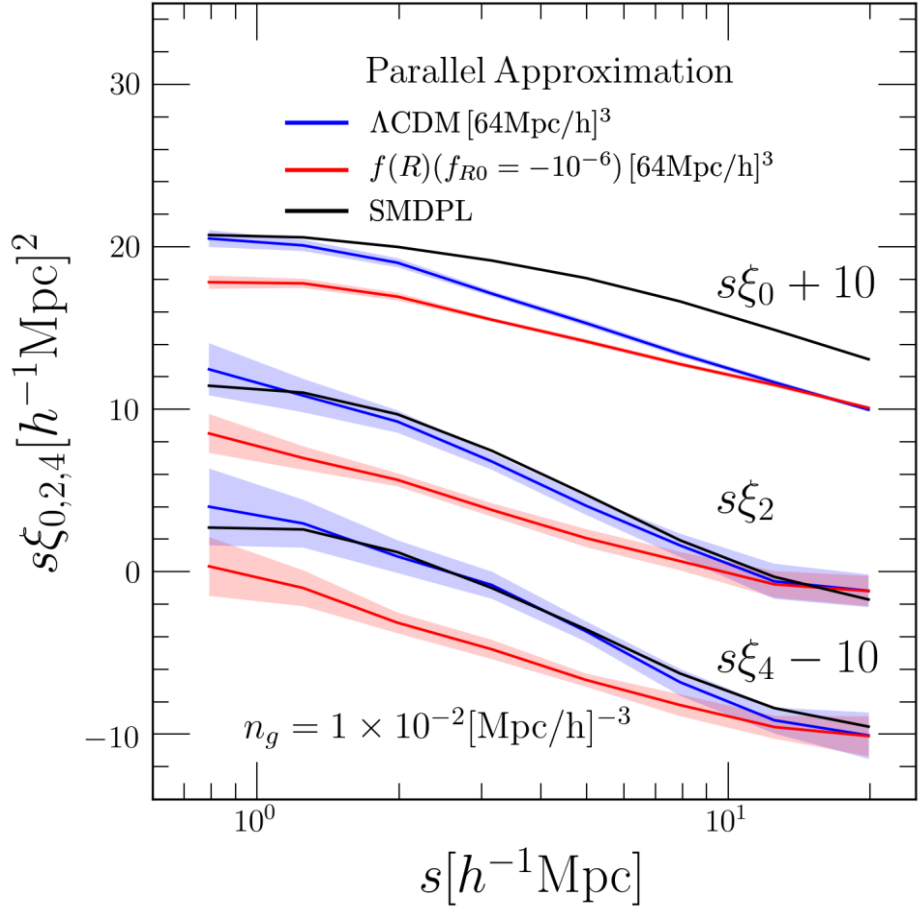

**Extended Data Figure 5. | The robustness of small-scale redshift-space clustering against the simulation box size.** Redshift-space multipoles from the SMDPL simulation used in the main paper (black line), are compared to the same statistics measured from a ΛCDM simulation (blue line) with the same initial conditions and box size as our state-of-the-art $f(R)$ simulation [35] (shown by the red line). The shaded regions give the $1\sigma$ uncertainty around the expectation value, derived from 500 realizations with line-of-sight along different directions of the simulation box. The comparison of the black and blue lines shows that the lack of long-wavelength modes due to the small box size has a strong impact on the monopole $\xi_0$, which shows a significant deficit of power for all scales larger than ~1 Mpc/h. Remarkably, however, the higher order multipoles from the two ΛCDM simulations are in very good agreement within the scatter, indicating a negligible impact on the quadrupole $\xi_2$ and hexadecapole $\xi_4$ of the missing long-wavelength modes. This is crucial for our conclusions, as these are the quantities containing most of the velocity-field information and, as such, providing the most robust test of gravity (see text in **Methods).**

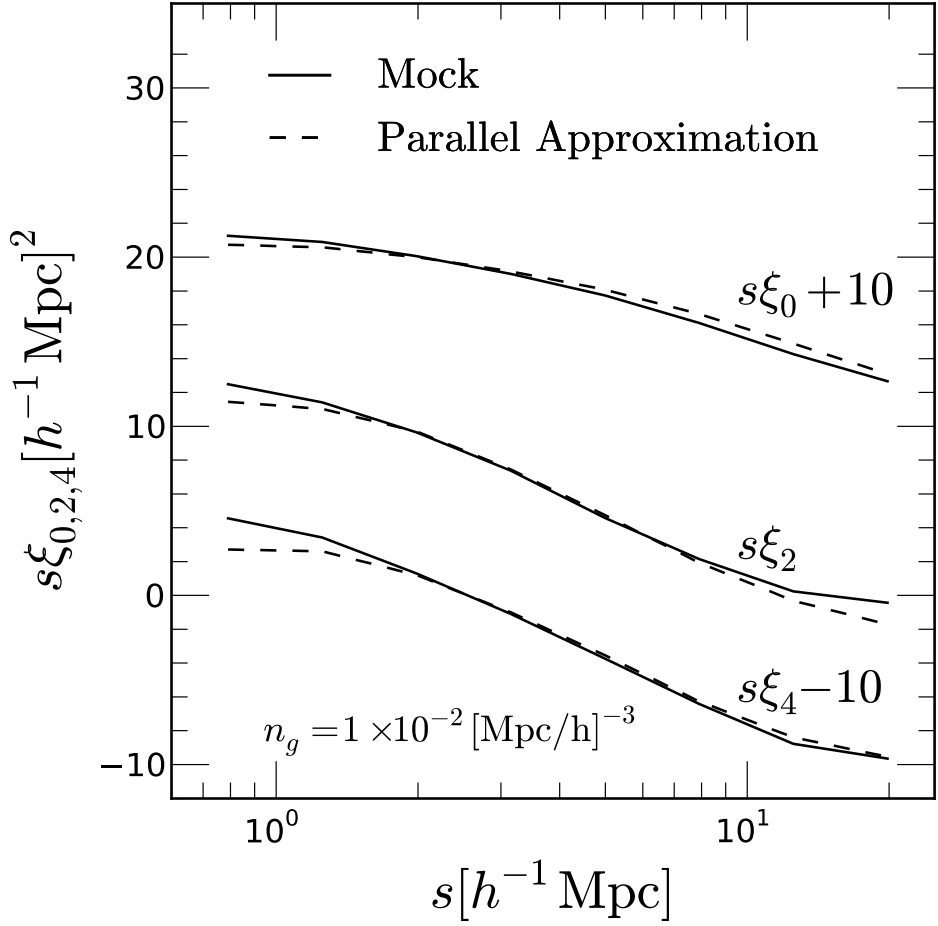

**Extended Data Figure 6. | The robustness of our tests to wide-angle effects.** The multipoles of the two-point correlation function $\xi(r_\sigma, r_\pi)$ are compared for the two cases in which the parallel approximation (dashed lines, see text) or a fully realistic SDSS-like mock survey are used. Although there are noticeable effects on the monopole $\xi_0$, over the range 2~10 Mpc/h both the quadrupole and the hexadecapole (i.e. the two most important statistics in our gravity test), show little difference between the two ways of implementing *redshift space* effects. This indicates the range where the parallel approximation, which had to be adopted for the $f(R)$ simulation (and for the tests of the two previous figures), can be safely compared with a fully realistic description as in the ΛCDM mock samples and, obviously, in the real data.